\documentclass[%
reprint,
groupedaddress,
preprintnumbers,
amsmath,amssymb,
aps,
longbibliography,
]{revtex4-1}

\usepackage{graphicx}
\usepackage{dcolumn}
\usepackage{bm}
\usepackage{hyperref}
\usepackage{color}

\begin{document}

\title{
Three-particle correlation from a Many-Body Perspective:\\Trions in a Carbon Nanotube
}

\author{Thorsten Deilmann}
  \email{thorsten.deilmann@uni-muenster.de}
  \affiliation{Institut f\"ur Festk\"orpertheorie, Westf\"alische Wilhelms-Universit\"at M\"unster, 48149 M\"unster, Germany.}%
\author{Matthias Dr\"uppel}
  \affiliation{Institut f\"ur Festk\"orpertheorie, Westf\"alische Wilhelms-Universit\"at M\"unster, 48149 M\"unster, Germany.}%
\author{Michael Rohlfing}
  \affiliation{Institut f\"ur Festk\"orpertheorie, Westf\"alische Wilhelms-Universit\"at M\"unster, 48149 M\"unster, Germany.}%

\date{4 May 2016}%

\begin{abstract}
Trion states of three correlated particles (e.g., two electrons and one hole)
are essential to understand the optical spectra of doped or gated nanostructures, like
carbon nanotubes or transition-metal dichalcogenides.
We develop a theoretical many-body description for such correlated states
using an ab-initio approach.
It can be regarded as an extension of the widely used $GW$ method and Bethe-Salpeter equation,
thus allowing for a direct comparison with excitons.
We apply this method to a semiconducting (8,0) carbon nanotube,
and find that the lowest optically active trions are red-shifted by $\sim 130$\,meV compared to the excitons,
confirming experimental findings for similar tubes. 
Moreover, our method provides detailed insights in the physical nature of trion states.
In the prototypical carbon nanotube we find a variety of different excitations, discuss 
the spectra, energy compositions, and correlated wave functions. 
\end{abstract}
\pacs{73.22.-f,78.67.Ch,71.15.Qe}

\maketitle

Trions or ``charged excitons'' compete with excitons in the
luminescence of carbon nanotubes (CNTs)
\cite{CNT_tr7,CNT_tr5,CNT_tr6,CNT_tr1,CNT_tr11,CNT_tr8,CNT_tr9,CNT_tr2,CNT_tr3,CNT_tr4,CNT_tr10}
and transition-metal dichalcogenides
\cite{MoS2_tr2,MoS2_tr1,MoS2_tr3,MoS2_tr4}.
They result from correlation between (light-activated) excitons and 
additional charges (from doping or gating).
However, detailed theoretical information about trions in atom-scaled
systems, like CNTs, is very difficult to achieve for conceptual and 
numerical reasons.
Here we develop an ab-initio many-body method  that can describe trions on top of the widely used $GW$ method and Bethe-Salpeter equation (BSE).
Our approach can thus be considered as an extension of the many-body perturbation theory (MBPT)  \cite{PhysRevB.62.4927,RevModPhys.74.601} to the situation of three correlated quasi-particles.

Trions are composed from electrons and holes forming correlated three-particle states in a condensed-matter system.
In this work, we concentrate on two electrons and one hole
(the treatment of a trion formed by two holes and one electron would be analogous and we expect similar results \cite{CNT_tr6}).
Usually one electron and one hole result from interband excitation by light,
like an exciton in a charge-neutral system, while the third particle stems from impurities, intentional doping, or can be induced by a gate voltage.
The actual source of the additional electron (doping, field effects, etc.)
is not considered for our present description of trions.
A trion state of two electrons and one hole with a total momentum {\bf K} can be described as
a linear combination of products of single-particle wave functions
$\phi_{n{\bf k}}(x)$,
\begin{align}
\label{eq01}
\Phi^{(T,{\bf K})}&(x_h, x_1, x_2) =
\sum_{{\bf v},{\bf c}_1,{\bf c}_2} 
A^{(T,{\bf K})}_{{\bf v},{\bf c}_1,{\bf c}_2}
\phi^\ast_{{\bf v}}(x_h)  \times\nonumber\\
&\times \frac{1}{\sqrt{2}} \left\lbrace
\phi_{{\bf c}_1}(x_1) \phi_{{\bf c}_2}(x_2) -
\phi_{{\bf c}_2}(x_1) \phi_{{\bf c}_1}(x_2) 
\right\rbrace
\end{align}
with ${\bf v}=(v,{\bf k}_v)$ denoting band index and wave number of a hole
in the valence bands (analogously, ${\bf c}_1$ and ${\bf c}_2$ for the two
electrons in the conduction bands) and $x_h$, $x_1$, $x_2$ denoting their
coordinates (position and spin).
Due to Bloch's theorem, the total momentum {\bf K} is a good quantum number, and
only wave vectors (from the first Brillouin zone) with
${\bf k}_1 + {\bf k}_2 - {\bf k}_v = {\bf K}$ contribute to the sum.
The trion states described within our framework (determined by the coefficients $A^{(T,{\bf K})}_{{\bf v},{\bf c}_1,{\bf c}_2}$)
are eigenstates of an effective Hamiltonian with matrix elements
\footnote{See Supplementary Material for a detailed discussion of our theoretical 
framework and further numerical details at \url{http://link.aps.org/supplemental/10.1103/PhysRevLett.116.196804},
which includes Ref.~\cite{tr_1,tr_3,tr_4,lda_pz,pp_ham,pp_kb,SLEPc2}.}
\begin{flalign}\label{eq2}
  \langle {\bf v}{\bf c}_1{\bf c}_2| \hat{H}^{(eeh)} &|{\bf v}'{\bf c}'_1{\bf c}'_2\rangle = &\\
&      (\epsilon_{{\bf c}_1} + \epsilon_{{\bf c}_2} - \epsilon_{\bf v}) 
\delta_{{\bf c}_1,{\bf c}_1'} \delta_{{\bf c}_2,{\bf c}_2'} 
\delta_{{\bf v},{\bf v}'} 
&(\hat{H}_{\text{BS}})\quad\nonumber\\
+ &  (W_{{\bf c}_1{\bf c}_2,{\bf c}'_1{\bf c}'_2} - 
     W_{{\bf c}_1{\bf c}_2,{\bf c}'_2{\bf c}'_1}) \delta_{{\bf v},{\bf v}'} 
&(\hat{H}_{ee})\quad\nonumber\\
- &   (W_{{\bf v}'{\bf c}_1,{\bf v}{\bf c}'_1} - 
      V_{{\bf v}'{\bf c}_1,{\bf c}'_1{\bf v}})
     \delta_{{\bf c}_2,{\bf c}_2'} 
&(\hat{H}_{eh,1})\quad\nonumber\\
- &   (W_{{\bf v}'{\bf c}_2,{\bf v}{\bf c}'_2} - 
      V_{{\bf v}'{\bf c}_2,{\bf c}'_2{\bf v}})
     \delta_{{\bf c}_1,{\bf c}_1'} 
~ .
&(\hat{H}_{eh,2})\quad\nonumber
\end{flalign}
Here, $\epsilon_{{\bf c}_1} \equiv \epsilon_{c_1,{\bf k}_1}$
etc. denote the band-structure energies of a preceding
$GW$ calculation.
The first line of Eq. (\ref{eq2}) describes
the single particle contributions given by the system's band 
structure, while the other terms describe the interaction
(direct and exchange) between the two electrons, and between the hole
and each of the electrons.
Without the second electron Eq.~(\ref{eq2}) is reduced to
$\langle {\bf v}{\bf c} | \hat{H}^{(eh)} | {\bf v}'{\bf c}' \rangle =
 (\epsilon_{\bf c}-\epsilon_{\bf v})\delta_{{\bf c}{\bf c}'}\delta_{{\bf v}{\bf v}'} - 
 (W_{{\bf v}'{\bf c},{\bf v}{\bf c}'} - V_{{\bf v}'{\bf c},{\bf c}'{\bf v}})$,
which are the matrix elements of the BSE Hamiltonian commonly used for excitons within MBPT \cite{Strinati82,Strinati84}.
Eq.~(\ref{eq2}) can be regarded as an extension of the BSE from two to three particles,
allowing to directly compare the spectra of trions (from Eq.~(\ref{eq2})) and excitons (from the BSE) 
in a consistent way.
In semiconductor quantum systems, similar Hamiltonians are commonly employed \cite{Esser2000,Filinov2004,Narvaez2005,Rabani2008,QP_tr4,QP_tr3} (see next paragraph).
A detailed discussion can be found in the Supplementary Material \cite{Note1}.

The general ab-initio determination of the Hamiltonian (\ref{eq2}) constitutes a
key issue of our study.
In semiconductor quantum dots and similar structures, in which the 
relevant length scales are much larger than interatomic distances, 
$\hat{H}^{(eeh)}$ may be constructed from
empirical parameters (e.g., effective masses, one single 
dielectric-constant value, etc.)
\cite{Esser2000,Filinov2004,Narvaez2005,Rabani2008,QP_tr4,QP_tr3}.
In a CNT, on the other hand, excitonic binding occurs on a much smaller
length scale of about one nanometer.
Inter-atomic electronic-structure details become important, and
dielectric screening properties are inhomogeneous and anisotropic on
such small length scale.
A priori, it is not clear if a
parameter-controlled modelling of $\hat{H}^{(eeh)}$ \cite{Esser2000,Filinov2004,Narvaez2005,Rabani2008,QP_tr4,QP_tr3} is sufficient.
To overcome system specific modelling,
we perform a general determination of $\hat{H}^{(eeh)}$
for one-, two-, and three-dimensional systems
(and gain full access to the atomic composition). This
must be performed
on a microscopic level, including many-body effects, starting from atom-
and orbital-resolved single-particle wave functions $\phi_{n{\bf k}}(x)$.
Thereby the six-dimensional dielectric function $\epsilon({\bf r},{\bf r}')$
evaluated from
$\phi_{n{\bf k}}(x)$ and their energies $\epsilon_{n{\bf k}}$ within 
the random-phase approximation is also included
[rather than using a dielectric constant $\epsilon$ or a simple distance
dependence $\epsilon(|{\bf r}-{\bf r}'|)$].
The dielectric function is needed for the screened Coulomb interaction
($W({\bf r},{\bf r}') = 
\int \epsilon^{-1}({\bf r},{\bf r}'') V({\bf r}'',{\bf r}') d^3r''$) which
contains the full information about spatial inhomogeneity and
anisotropy.
Such procedure was the key to understand excitons in CNTs
\cite{CNT_Louie,Chang2004,CNT_impurity,CNT_GdW}
within an ab-initio $GW$/BSE approach.
We note that (in contrast to many model Hamiltonians)
the screened interaction ($W_{{\bf c}_1{\bf c}_2,{\bf c}_1'{\bf c}_2'}$ etc.)
within $GW$/BSE (see e.g. \cite{Johnson87,Puschnig12,Milko12,Milko13,MetalMolLouie,Garcia2009,Garcia2011,CNT_GdW})
acts in a natural way on the band structure ($\epsilon_{\bf{v}}$ etc.),
on the excitonic binding, and on the trions on equal footing.
We also note that polaronic or self-trapping effects, localized charges at defects/dopant atoms,
as well as high-density effects (bleaching, band-gap renormalization etc.) are not considered within our framework.

In contract to excitons (described by the BSE),
the configurational space of trions is much larger,
which prohibits the standard diagonalization of the
Hamiltonian of Eq.~(\ref{eq2}).
For instance, our typical calculations for the (8,0) nanotube
include 16 valence bands, 16 conduction bands, and 32 $\bf k$-points,
yielding a trion configuration space of
$\sim \frac{1}{2} 16^3\times32^2$ $\sim$ $2.1\times 10^6$
(as compared to an exciton configuration space of only
$16^2\times32$ = 8192).
The unit cell contains $32$ atoms (Fig.~S1 of the Supplementary Material),
for the evaluation of excitons and trions
this cell is extended by a factor of 32 (due to the usage of 32 $\bf k$-points).
This extended unit cell is shown in Fig.~\ref{fig_3}.
Fortunately, all Coulomb interaction terms (bare or screened, direct or
exchange) in Eq. (\ref{eq2}) {\em are of two-particle nature only},
and the resulting Hamilton matrix is very sparse \cite{Note1}.
This opens an avenue to deal with the trion Hamiltonian iteratively, either
by Haydock recursion or similar \cite{haydock,timeint} when focusing on spectra, or by
the Lanczos algorithm or similar \cite{SLEPc1} when eigenstates are requested.
Such an iterative procedure, together with parallel architecture
make our approach feasible \footnote{%
Sufficient processors and memory
were supplied by the HPC system PALMA of the University of M{\"u}nster
and the super-computer JURECA at J{\"u}lich Supercomputing Centre (JSC).}.

\begin{figure}
  \includegraphics[scale=0.9]{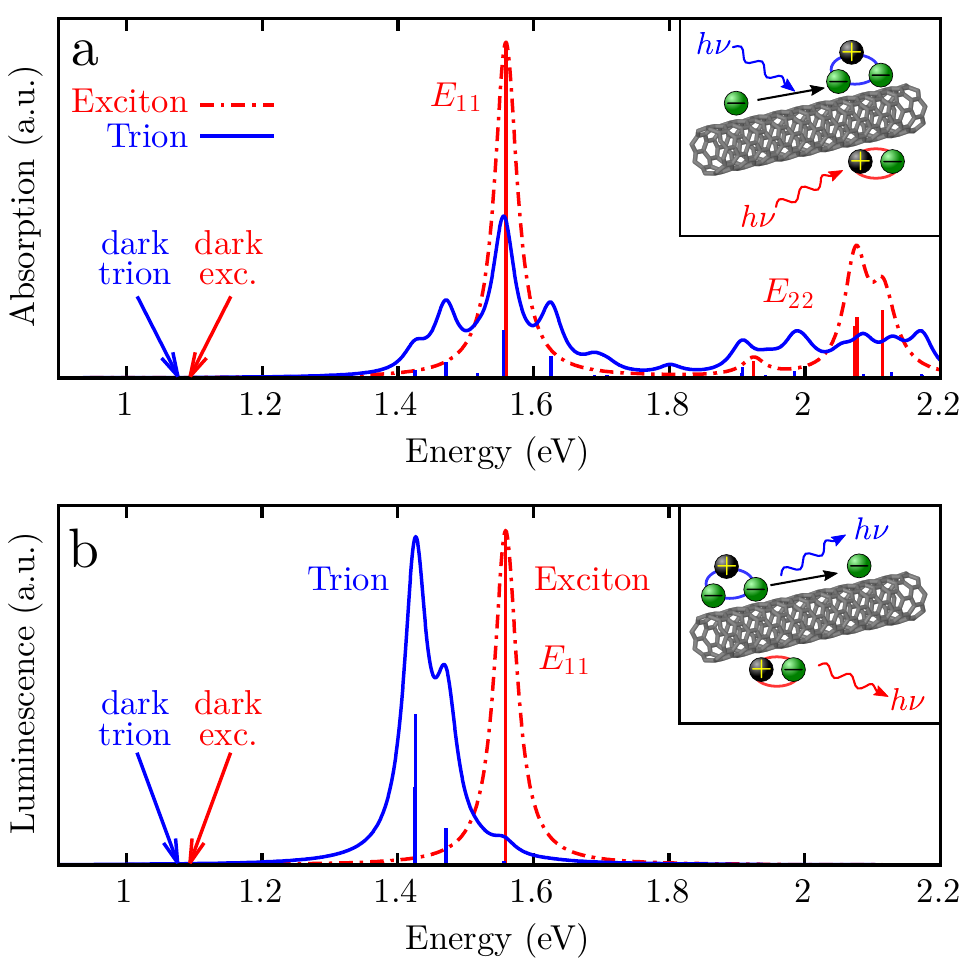}
\caption{
Optical absorption and luminescence spectrum of an (8,0) 
carbon nanotube.
(a) Optical absorption due to excitons 
(red dashed lines) and due to negatively charge trions (blue lines).
The arrows at $1.09$\,eV (exciton) and at 
$1.07$\,eV (trion) indicate the onset of the spectrum, which is
not visible because of zero optical dipole strength of the 
corresponding states.
(b) Luminescence spectrum from the
$E_{11}$ exciton and from the related optically allowed trions, assuming 
that the occupancy of the excited states relative to each other is
given by a Boltzmann distribution at room temperature
(see Supplementary Material).
This distribution highlights the
lower-energy states near $1.4$\,eV and suppresses states above $1.5$\,eV.
In all cases the electric-field vector of the light is
along the tube axis.
Note that the amplitudes of the exciton and trion spectrum cannot
be compared with one another because the trion spectrum scales with
the density of donated electrons.
The trions have been calculated for ${\bf K}=0$, at which the optical 
dipoles are by far the strongest. We consider transitions to the conduction-band minimum.
Including non-zero {\bf K} yields very similar results.
An artificial broadening of $0.02$\,eV is used.
}\label{fig_1}
\end{figure}

Having set up a general method to describe trions,
we are now able to discuss trions in a prototypical
low-dimensional system on the atomic scale, i.e. the (8,0) CNT.
Figure~\ref{fig_1} shows our calculated absorption spectra of excitons and trions 
(upper panel) and corresponding room-temperature luminescence spectra (lower panel).
We want to stress that in most common experimental set-ups
the spectra show the combined effects of trions and excitons.
The exciton spectrum exhibits two distinct peaks ($E_{11}$ and $E_{22}$),
but there are also many dark excitons (triplet excitons and
dipole-forbidden singlet excitons), starting at an onset of $1.09$\,eV
\footnote{%
Note that due to the involved approximations,
the \textit{absolute} accuracy of any $GW$/BSE study cannot be better than $\sim$0.1\,eV.
On the other hand, \textit{relative} energy differences between excitations
and between excitons and trions can be given with much higher precision.
}.
These excitons are accompanied by a variety of
trion states.
At the low-energy onset, the first trion state is found at $1.07$\,eV.
The first {\em dipole-allowed} trion, however, is found slightly below
the $E_{11}$ exciton, at $1.43$\,eV, together with several other dipole-allowed trions
below and above the $E_{11}$ exciton.
More states are observed at and above $1.8$\,eV.
A more drastic picture emerges in the luminescence spectrum, which contains the same 
excitations, but with weights that simulate a luminescence experiment \cite{Note1,Note4}.
Here we simply assume that the trions (and excitons as well) achieve
thermal equilibrium after the excitation process, with energy dependent occupation probabilities given by a 
room-temperature Boltzmann distribution (see Supplementary Material). 
Luminescence from trions then occurs
predominantly from those states near $1.4$\,eV, i.e. $\sim$130\,meV below
the $E_{11}$ exciton,
which dominates the exciton luminescence
\footnote{Note that the dynamical processes leading to thermal equilibrium,
as well as details of the emission processes,
are not considered here,
so the amplitudes in the luminescence spectrum can only indicate trends.
}.

Experimental spectra similar to Fig.~\ref{fig_1} 
have been measured for various CNTs, with red-shifts of 
100-200\,meV \cite{CNT_tr7,CNT_tr6,CNT_tr1} between trion and exciton luminescence.
Unfortunately, we are not aware of any trion experiment on an (8,0) CNT yet,
and most CNTs from experiment have more complicated chirality and 
are thus numerically too demanding for our present study.
However, our results show redshifts of the same size.
We note that the trionic binding effects between exciton and electron
strongly differ from state to state.
For example, the dark states near $1.1$\,eV exhibit a much smaller redshift of only $\sim$20\,meV.

\begin{table}[t]%
\centering
\caption{
Energy composition of the exciton and trion energies.
Transition energy $\Omega$, band-structure term 
$\langle E_\text{BS} \rangle$, 
electron-hole interaction $\langle E_{eh} \rangle$,
and electron-electron interaction $\langle E_{ee} \rangle$
for the lowest dark and bright states (in eV), for zero total momentum
(exciton: ${\bf Q}=0$, trion: ${\bf K}=0$).
For further details see main text.
}\label{tab1}
  \begin{ruledtabular}
  \begin{tabular}{cccccc}
 (eV)  & $\Omega$ & $\langle E_\text{BS} \rangle$ & $\langle E_{eh} \rangle$ & $\langle E_{ee} \rangle$ \\

  \hline
  dark exciton   & $1.09$ & $2.17$ & $-1.08$ & -- \\
  dark trion     & $1.07$ & $2.18$ & $-1.65$ & $0.56$ \\
  \hline
  bright exciton & $1.56$ & $2.51$ & $-0.95$ & -- \\
  bright trion   & $1.43$ & $2.60$ & $-1.69$ & $0.52$ \\
  \end{tabular}
  \end{ruledtabular}
\end{table}%

\begin{figure}
  \includegraphics[width=\linewidth]{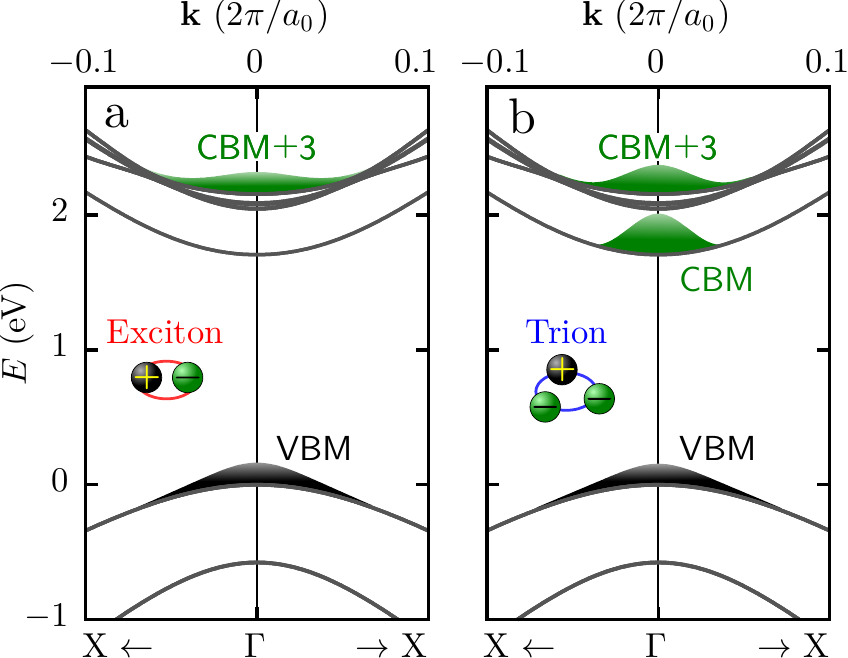}
\caption{
Band structure of an (8,0) CNT, showing the composition of the 
lowest bright exciton (a) and trion (b).
The black bell-shaped distributions on the highest valence band indicate the
composition of the hole as a function of wave number ${\bf k}_v$ (resulting from
the coefficients of the exciton and the trion, respectively).
Similarly, the green bell-shaped distributions on the lowest conduction band
and the CBM+3 band indicate the contributions to the electrons (see text).
The total momenta of the exciton and trion were chosen as zero.
}\label{fig_2}
\end{figure}

\begin{figure*}
  \includegraphics{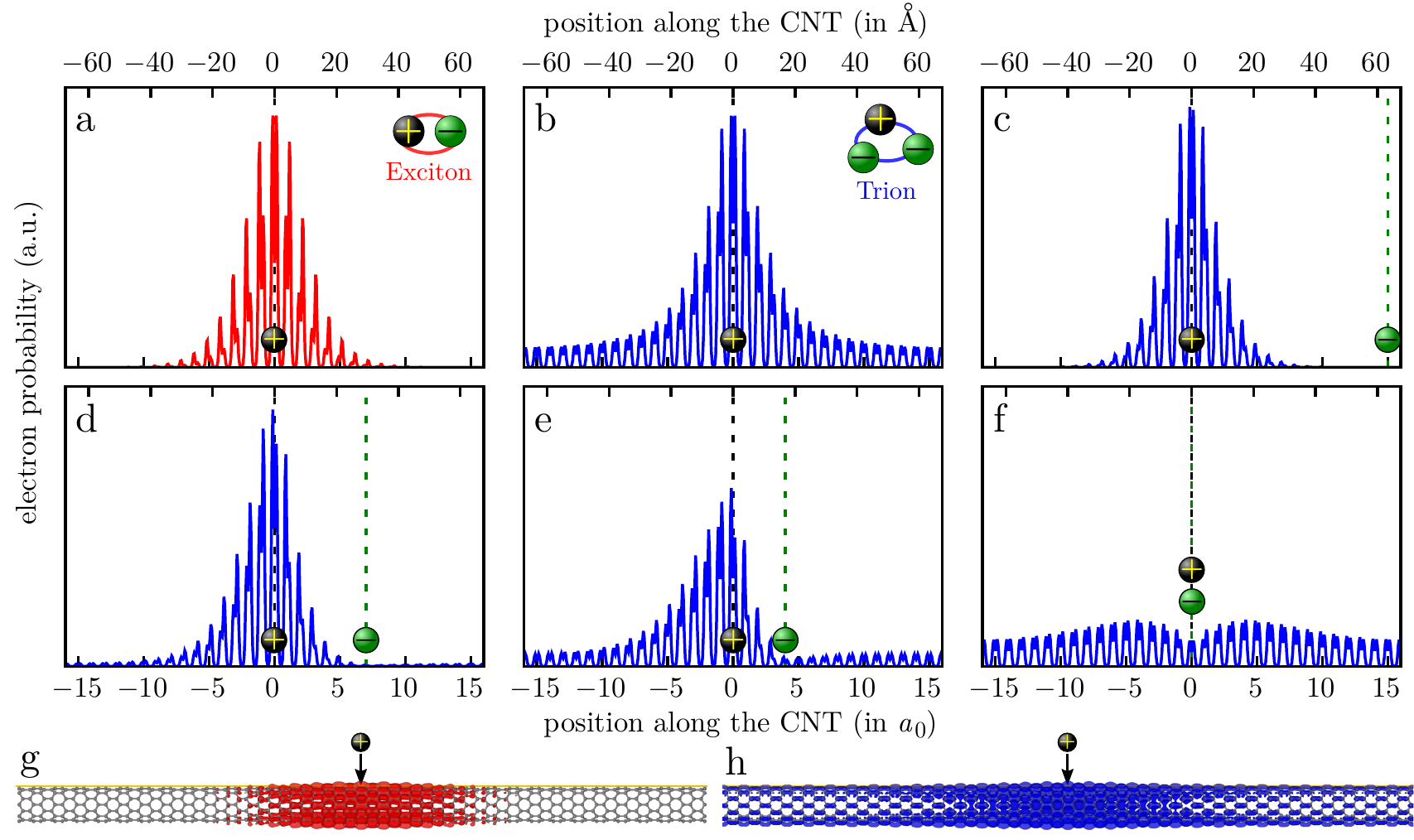}
\caption{
Spatial electron distribution in the exciton (red) and trion (blue) as a function of position
along the CNT.
(a) Spatial distribution of the electron relative to the hole, which is kept
at the center of the panel (black vertical line), for the lowest-energy exciton.
(b) Distribution of one electron relative to the hole (again in the center)
for the lowest-energy trion, after complete spatial averaging over the other
electron.
(c)-(f) Distribution of one electron of the lowest-energy trion.
The hole and the other electron are kept in the center (black vertical line) and
at positions (indicated by the green vertical line) of 15, 7, 4, and 0 lattice 
constants $a_0$ (63\,\AA, 30\,\AA, 17\,\AA, and 0\,\AA) away from the hole.
All distribution functions are given by $|\Phi|^2$ of the wave functions of 
Eq. (\ref{eq01}) and analog for excitons \cite{Note1}.
All positions are on a line parallel to the CNT, at a distance of $3.7$\,\AA\
from its axis (i.e. on the outside of the CNT, $0.55$\,\AA\ above the nuclei, 
see also yellow line in (g) and (f)).
(g)-(h) Three-dimensional view of the electron distribution, see (a) and (b).
}\label{fig_3}
\end{figure*}

We now discuss the energetics of the trions in comparison to excitons (Tab.~\ref{tab1}).
For an exciton, $\Omega$ is the eigenvalue of the corresponding Hamiltonian.
For a trion, $\Omega$ is given as the difference between the eigenvalue of the
Hamiltonian (\ref{eq2}) and the CBM energy (see Supplementary Material), which allows a systematic comparison.
$\Omega$ thus immediately gives the transition energy for the
transition from the trion to the electron remaining in the CBM at zero momentum
(i.e., at the $\Gamma$ point).
For the trions, $\langle E_\text{BS} \rangle$ is also
given relative to the CBM. This band is chosen because the (radiative) decay of a trion (with ${\bf K}=0$) is mostly 
into the CBM, so the calibration to the CBM immediately yields the 
luminescence frequency of the transition (cf. Fig.~\ref{fig_1}).
All other values are expectation values of the corresponding terms to the
Hamiltonians $\langle E_i \rangle = \langle \Phi | \hat{H}_{i} | \Phi \rangle$,
with $\Phi$ denoting the wave function.\\
Both for excitons and for trions we discuss in Tab. \ref{tab1} the 
lowest-energy states (which happen to have zero dipole moment, i.e. 
these are dark states) and the first bright states.
The compositions of the bright states from bands and wave numbers 
is also shown in Fig.~\ref{fig_2}.
Both dark states and both bright states show about the same band-structure contribution,
$\langle E_\text{BS} \rangle$, indicating similar composition from the bands.
The bright exciton, e.g., is composed from the highest valence band and from
the CBM+3 conduction band (excitons from lower conduction bands are dark,
or involve deeper valence electrons which leads to higher transition energies).
The bright trion similarly involves one electron in the CBM+3 band (see Fig.~\ref{fig_2}) while
the other electron is mostly located in the CBM band.
The main difference between excitons and trions is in the Coulomb interaction.
The trions observe electron-electron repulsion $\langle E_{ee} \rangle$ 
which is of course absent in the excitons.
They also observe stronger electron-hole attraction because there are
{\em two} electrons to which the hole is attracted.
However, $\langle E_{eh} \rangle$ is {\em less than twice} as strong compared to
the exciton, because the electron-electron repulsion drives the electrons somewhat
apart from each other, thus weakening the attraction of both of them to the hole.

Detailed information about spatial correlation can be obtained from
the wave function (\ref{eq01}).
Here we focus on the dark lowest-energy states.
The wave function of the first bright exciton and trion look similar.
Because of the six-/nine-dimensional nature of the exciton/trion wave functions,
we discuss some characteristic features, only.
Figs.~\ref{fig_3} a and g show the probability distribution of the electron in the exciton state
along the CNT, relative to the hole which is fixed in the center of the panel.
Figs.~\ref{fig_3} b and h exhibit the same quantity for the trion, after averaging over the
coordinates of one of the two electrons.
Figs.~\ref{fig_3} b and h thus show the correlation of any of the two electrons to the hole.
Due to the electron-hole attraction,
the localization around the hole is still quite strong
but farther extended (in real space) than for the exciton,
resulting from the electron-electron repulsion.
In particular, the envelope function does not decay to zero at large distance, but
converges to the CBM wave function: at large distance, one
electron observes attraction to the hole and repulsion from the other electron
simultaneously, which cancel such that some free-electron mobility remains.
The larger spatial extent of the electrons corresponds to stronger localization
in reciprocal space as compared to the electron in the exciton state 
(see Fig.~\ref{fig_2}).
In contrast, the reciprocal-space distribution of the {\em hole} is nearly the
same for exciton and trion.

Further details of the trion become visible in Figs.~\ref{fig_3} c-f where the hole 
and one electron are kept fixed (marked by a black line in the center and by a 
green line at various distances from the center of each panel), showing
the spatial distribution of the other electron.
In panel c the first electron is so far away from the hole that the same
electron-hole pair distribution as for the exciton emerges.
In panel d the second electron observes repulsion from the first electron (which is
$30$\,\AA\ away from the hole) and tends to swerve to the left.
This becomes even more pronounced when the fixed electron approaches the hole further 
(panel e). Here their distance of $17$\,\AA\ 
is already smaller than the intrinsic size of the exciton, so the above-mentioned
free-particle behaviour of the other electron becomes visible.
At even closer electron-hole distance (panel f), the other electron behaves as a free
particle which is attractively scattered at the electron-hole pair.
Note that in all cases the Pauli principle suppresses the amplitude of the
two electrons at the same position (if they have the same spin), resulting e.g. 
in the dip in the center of panel f.

In summary, we have introduced a novel ab-initio scheme that accounts for many-body effects,
which enables us to accurately describe trions and directly compare them to excitons. 
It further allows studying in detail the mechanisms of inter-particle
correlation and entanglement on the atomic level.
We have applied this method to negatively charged excitons (trions) in a
semiconducting (8,0) CNT.
A variety of trion states with different binding energies are found,
e.g., $\sim$20\,meV for the lowest dark and $\sim$130\,meV for the lowest bright trion.
The latter dominates luminescence which is red-shifted compared to the excitons,
confirming experimental observations.
Our data allows a detailed description of the internal energetics of trion
states, their correlated composition from the single-particle states,
and their spatial wave functions on a microscopic level.
The latter demonstrate how quantum-mechanical correlation works on the
atomic scale in nanostructured condensed matter.
The presented method is generally applicable to all nanostructures
and the obtained results are a further step towards a deeper understanding 
of three particle correlation and towards a specific manipulation of the latter by light and by charging,
e.g., due to an externally applied gate voltage.

We thank P. Kr\"uger for fruitful discussions and for critically reading our
manuscript.
The authors gratefully acknowledge the
computing time granted by the John von Neumann
Institute for Computing (NIC) and provided on the 
super-computer JURECA at J\"ulich Supercomputing Centre (JSC).

%

\pagebreak
\begin{widetext}\begin{center}\bf\large%
Supplement to\\
Three-particle correlation from a Many-Body Perspective:\\Trions in a Carbon Nanotube
\end{center}\end{widetext}
\setcounter{equation}{0}
\setcounter{figure}{0}
\setcounter{table}{0}
\setcounter{page}{1}
\makeatletter
\renewcommand{\theequation}{S\arabic{equation}}
\renewcommand{\thefigure}{S\arabic{figure}}

\section{Many-body approach towards three correlated particles}

Ab-initio many-body perturbation theory (MBPT) is a state of the art method
to reliably describe excited electronic states in electronic-structure theory \cite{RevModPhys.74.601,PhysRevB.62.4927}. 
Here we first briefly outline MBPT for electron-hole pairs for comparison and to provide
a common notation.
Thereafter corresponding issues for two electrons are outlined, followed by the
discussion of trionic states for two electrons and one hole. 
The discussion is carried out in second quantization and kept very brief and simple, for
explanatory purpose. Before deriving the Hamilton matrices in our many-body approach, Hartree-Fock theory is used for explanatory reasons only.

The formal description of correlated particles follows
general concepts \cite{tr_1,tr_3,tr_4} that have 
been employed for, e.g., semiconductor quantum dots  
\cite{Esser2000,Filinov2004,Narvaez2005,Rabani2008,QP_tr4,QP_tr3}.
Here we present the following derivations and formulas
to have a consistent notation to extend it.
Our many-body approach adds two essential features to those approaches:
(i) We incorporate inhomogeneous and anisotropic dielectric screening 
$\epsilon({\bf r},{\bf r}')$ on the atomic scale,
rather than using a dielectric constant $\epsilon$ or a simple distance 
dependence $\epsilon(|{\bf r}-{\bf r}'|)$.
(ii) The band-structure energies and the screened Coulomb interaction between
the particles both result from the same $GW$ self-energy operator (which
incorporates dielectric screening) and must be evaluated on equal footing.

\subsection{Electron-hole pair states}
\label{subsec_bse}

Within second-quantization notation, the ground state for an $N$-electron system
may be denoted as $|0\rangle$.
The application of annihilation and creation operators generates an electron-hole
pair configuration as
\begin{equation}
\label{supp_eq1}
|{\bf v}{\bf c}\rangle := \hat{a}^\dagger_{\bf c} \hat{a}_{\bf v} |0\rangle
~ .
\end{equation}
In here, ${\bf v}$ (${\bf c}$) denotes occupied valence (empty conduction) 
states. For a periodic system
${\bf v}=(v,{\bf k}_v$) and ${\bf c}=(c,{\bf k}_c$) combine the band index
($v$, $c$) with a wave number (${\bf k}_v$, ${\bf k}_c$) for holes and electrons.
Here we prefer the short-hand notation ${\bf v}$ (${\bf c}$) for the sake of brevity.

Within Hartree-Fock theory, the single-particle states $\phi_{\bf n}(x)$ are considered to
be the solutions of the Hartree-Fock equations (i.e. $|0\rangle$ is taken as a single
Slater determinant and minimizes the total energy of the system).
The many-body Hamiltonian is then given as
\begin{equation}
\label{supp_eq2}
   \hat{H} = \sum_{{\bf ij}} h_{{\bf ij}} \hat{a}^\dagger_{\bf i} \hat{a}_{\bf j}
   + \frac{1}{2} \sum_{{\bf ijmn}} V_{{\bf ij},{\bf mn}}~
   \hat{a}^\dagger_{\bf i} \hat{a}^\dagger_{\bf j} \hat{a}_{\bf n} \hat{a}_{\bf m}
\end{equation}
with $h_{\bf ij}$ being the matrix elements of the single-particle terms (kinetic energy and
external potential, e.g. pseudopotentials from the cores) and $V_{{\bf ij},{\bf mn}}$ being
matrix elements of the Coulomb interaction:
\begin{equation}
\label{supp_eq3}
V_{{\bf ij},{\bf mn}} = \int \phi_{\bf i}^\ast(x) \phi_{\bf j}^\ast(y) V(x,y) \phi_{\bf m}(x) \phi_{\bf n}(y) dx dy
   ~ .
\end{equation}
The coordinate $x=({\bf r},\sigma)$ combines position and spin; correspondingly,
the integration includes spin summation, i.e. 
$\int dx = \sum_{\sigma=\pm} \int d^3r$.
At the moment, the bare Coulomb interaction is considered, i.e.
$V(x,x') = e^2/|{\bf r}-{\bf r}'|$. The implementation of
screening effects (replacing $V$ by $W$) is discussed further below.

It turns out that the free electron-hole pair configurations $|{\bf v}{\bf c}\rangle$ are not
eigenstates of the Hamiltonian (\ref{supp_eq2}), because the matrix elements between
two configurations $|{\bf v}{\bf c}\rangle$ and $|{\bf v}'{\bf c}'\rangle$ are non-diagonal:
\begin{align}
\label{supp_eq4}
\langle {\bf v}{\bf c} | \hat{H}
|{\bf v}'{\bf c}'\rangle   =  &
(E_0^{({\rm HF})} + \epsilon^{({\rm HF})}_{\bf c} - \epsilon^{({\rm HF})}_{\bf v}) 
\delta_{{\bf c}{\bf c}'} 
\delta_{{\bf v}{\bf v}'}\nonumber\\
&- (V_{{\bf v}'{\bf c},{\bf v}{\bf c}'} - 
V_{{\bf v}'{\bf c},{\bf c}'{\bf v}})
\end{align}
with $\epsilon^{({\rm HF})}_n$ being the (Hartree-Fock) band-structure energy of level $n$.
The Hartree-Fock ground-state energy $E_0^{({\rm HF})}$ can be disregarded in the following.
Excited states result as linear combinations of the free
electron-hole pair configurations, given by the diagonalization of the
Hamilton matrix of Eq. (\ref{supp_eq4}):
\begin{equation}
\label{supp_eq5}
|S,{\bf Q}\rangle = \sum_{{\bf v}{\bf c}} B^{(S,{\bf Q})}_{{\bf v}{\bf c}} |{\bf v}{\bf c}\rangle
\end{equation}
with the coefficients $B^{(S,{\bf Q})}_{{\bf v}{\bf c}}$ and the excitation energy 
$\Omega^{(S,{\bf Q})}$ of
excitation $|S,{\bf Q}\rangle$ resulting from the eigenvalue equation
\vspace*{-.2cm}
\begin{equation}
\label{supp_eq6}
\sum_{{\bf v}'{\bf c}'} \langle {\bf v}{\bf c} | \hat{H} | {\bf v}'{\bf c}' \rangle B^{(S,{\bf Q})}_{{\bf v}'{\bf c}'} = \Omega^{(S,{\bf Q})} B^{(S,{\bf Q})}_{{\bf v}{\bf c}}
~ .
\end{equation}
These coupled states are the true electron-hole excitations (i.e., excitons in
periodic systems).
Note that the discussion here is restricted to the Tamm-Dancoff approximation
within time-dependent Hartree-Fock theory.
Eqs. (\ref{supp_eq5}) and (\ref{supp_eq6}) contain the total momentum {\bf Q} of
the exciton, which (according to Bloch's theorem) is a good quantum number of the
exciton and can be preselected.
Accordingly, the summations in Eqs. (\ref{supp_eq5}) and (\ref{supp_eq6}), which
contain double summations over ${\bf k}_v$ and ${\bf k}_c$, are restricted to 
such ${\bf k}_v$ that fulfill ${\bf k}_v={\bf k}_c-{\bf Q}$.

The electron-hole pair configurations yield a real-space amplitude of
\begin{equation}
\label{supp_eq7}
\Phi_{{\bf v}{\bf c}}(x_h,x_e) = 
\langle 0 | \hat{\psi}^\dagger(x_h) \hat{\psi}(x_e) |{\bf v}{\bf c}\rangle
= \phi_{\bf v}^\ast(x_h) \phi_{\bf c}(x_e) \hspace*{-.1cm}
\end{equation}
(employing annihilation and creation field operators, with $x_h$ and 
$x_e$ denoting the coordinates of the hole and the electron),
with a corresponding linear combination for a coupled state 
$|S,{\bf Q}\rangle$:
\begin{equation}
\label{supp_eq8}
\Phi^{(S,{\bf Q})}(x_h,x_e) = \sum_{{\bf v}{\bf c}} B^{(S,{\bf Q})}_{{\bf v}{\bf c}} \phi_{\bf v}^\ast(x_h) \phi_{\bf c}(x_e)
\end{equation}
(again with the restriction ${\bf k}_v={\bf k}_c-{\bf Q}$ in the 
summation).

Furthermore, optical dipole transitions between the ground state and an excited state
are controlled by the dipole-moment operator,
$\hat{{\bf p}} = \sum_{\bf ij} {\bf p}_{\bf ij} \hat{a}^\dagger_{\bf i} \hat{a}_{\bf j}$, yielding
\begin{eqnarray}
\label{supp_eq7a}
\langle 0 | \hat{{\bf p}} |{\bf v}{\bf c}\rangle & = & {\bf p}_{{\bf v}{\bf c}} \\
\label{supp_eq7b}
\langle 0 | \hat{{\bf p}} |{S,{\bf Q}}\rangle & = & \sum_{{\bf v}{\bf c}} B^{(S,{\bf Q})}_{{\bf v}{\bf c}} {\bf p}_{{\bf v}{\bf c}}
~ .
\end{eqnarray}
Note that only states with ${\bf Q}=0$ yield a non-zero dipole moment.

A formula similar to Eq.~(\ref{supp_eq4}) can be derived independently within many-body perturbation theory.
However, two important changes result from the consideration of dielectric screening
effects: (i) the Hartree-Fock energy levels are replaced by quasiparticle levels
$\epsilon_{\bf n}$
from a $GW$ calculation, and (ii) the interaction matrix elements between the
electron-hole pair configurations becomes screened in the direct interaction part.
We thus obtain an effective Hamiltonian for electron-hole pair states:
\begin{equation}
\label{supp_eq9}
\langle {\bf v}{\bf c}| \hat{H}^{(eh)}
|{\bf v}'{\bf c}'\rangle   =  
(\epsilon_{{\bf c}} - \epsilon_{\bf v}) 
\delta_{{\bf c}{\bf c}'}
\delta_{{\bf {\bf v}}{\bf {\bf v}}'}
- (W_{{\bf v}'{\bf c},{\bf v}{\bf c}'} - 
V_{{\bf v}'{\bf c},{\bf c}'{\bf v}})
\end{equation}
which is employed in the well established $GW$/BSE method.
The matrix elements $W_{\bf ij,mn}$ have the same structure as Eq.~(\ref{supp_eq3}),
but with the screened Coulomb interaction $W(x,x')$ which includes dielectric
screening effects in terms of the (inverse) dielectric function of the system
via $W({\bf r},{\bf r}') =
\int \epsilon^{-1}({\bf r},{\bf r}'') V({\bf r}'',{\bf r}') d^3r''$.
As in most $GW$/BSE studies, we use dynamic screening in the $GW$,
but restrict ourselves to static screening in the BSE.
Note that within MBPT the exchange interaction between electrons and holes results from
the classical part of the (bare) Coulomb interaction and is therefore not screened.

\subsection{Correlated states of two electrons}
\label{subsec_ee}

As a preparation for trion states, we take an intermediate step and discuss states
in which two electrons are moving in the conduction bands.
An independent-particle configuration would be given by
\begin{equation}
\label{supp_eq11}
|{\bf c}_1{\bf c}_2\rangle := \hat{a}^\dagger_{{\bf c}_2} \hat{a}^\dagger_{{\bf c}_1} |0\rangle
~ .
\end{equation}
To avoid double counting, we require that ${\bf c}_1 < {\bf c}_2$. This assumes that the
empty states are ordered. 
The definition of the order is arbitrary
(in particular when ${\bf c}=(c,{\bf k}_c$) includes a wave number) 
but has to be kept.

Within Hartree-Fock theory, the configurations (\ref{supp_eq11}) are again
not eigenstates of the Hamiltonian (\ref{supp_eq2}). In fact, the Hamilton
matrix elements between such configurations result as
\begin{align}
\label{supp_eq12}
\langle {\bf c}_1{\bf c}_2| \hat{H}
|{\bf c}_1'{\bf c}_2' \rangle   =  &
(E_0^{({\rm HF})} + \epsilon^{({\rm HF})}_{{\bf c}_1} + \epsilon^{({\rm HF})}_{{\bf c}_2}) 
\delta_{{\bf c}_1{\bf c}_1'}
\delta_{{\bf c}_2{\bf c}_2'}\nonumber\\
&+ (V_{{\bf c}_1{\bf c}_2,{\bf c}_1'{\bf c}_2'} - V_{{\bf c}_1{\bf c}_2,{\bf c}_2'{\bf c}_1'})
~ .
\end{align}
The diagonalization of Eq.~(\ref{supp_eq12}) yields correlated
electron-electron pair states.

The two-electron states (\ref{supp_eq11}) have a real-space amplitude of
\begin{align}
\label{supp_eq13}
\Phi_{{\bf c}_1{\bf c}_2}(x_1,x_2) &= \frac{1}{\sqrt{2}} \langle 0 | \hat{\psi}(x_1) \hat{\psi}(x_2) |{\bf c}_1{\bf c}_2\rangle\nonumber\\
&= \frac{1}{\sqrt{2}} \lbrace\phi_{{\bf c}_1}(x_1) \phi_{{\bf c}_2}(x_2) - \phi_{{\bf c}_2}(x_1) \phi_{{\bf c}_1}(x_2)\rbrace 
\end{align}
(with $x_1$ and $x_2$ being the coordinates of the two electrons)
which reflects the anti-symmetry required from two identical particles
(including the vanishing of the wave function for $x_1 = x_2$, i.e. same position and spin).

For physical correspondence to Eq.~(\ref{supp_eq9}) we again replace the Hartree-Fock levels
by QP energy levels (also stemming from the $GW$ approach) and screen the
interaction.
We thus obtain an effective Hamiltonian for electron-electron pair states:
\begin{align}
\label{supp_eq14}
\langle {\bf c}_1{\bf c}_2| \hat{H}^{(ee)}
|{\bf c}_1'{\bf c}_2' \rangle   =  &
(\epsilon_{{\bf c}_1} + \epsilon_{{\bf c}_2}) 
\delta_{{\bf c}_1{\bf c}_1'}
\delta_{{\bf c}_2{\bf c}_2'}\nonumber\\
&+ (W_{{\bf c}_1{\bf c}_2,{\bf c}_1'{\bf c}_2'} - W_{{\bf c}_1{\bf c}_2,{\bf c}_2'{\bf c}_1'})
~ .
\end{align}
Note that both interaction terms are screened. This is necessary to guarantee
invariance of the entire approach when the order of the single-particle
states (see at the beginning of this subsection) is selected in a different 
way.

\subsection{Trion states of two electrons and one hole}
\label{subsec_eeh}

As a direct extension of Secs. \ref{subsec_bse} and \ref{subsec_ee},
independent-particle configurations of two electrons and one hole are constructed as
\begin{equation}
\label{supp_eq21}
|{\bf v}{\bf c}_1{\bf c}_2\rangle := \hat{a}^\dagger_{{\bf c}_2} \hat{a}^\dagger_{{\bf c}_1} \hat{a}_{\bf v} |0\rangle
\end{equation}
with the restriction again that ${\bf c}_1 < {\bf c}_2$ within the chosen order of the 
single-particle levels, to avoid double counting among the identical electrons.

Within the framework of Hartree-Fock theory, the matrix elements between these
configurations are
\begin{align}
\label{supp_eq22}
\langle {\bf v}{\bf c}_1{\bf c}_2| \hat{H} &
	|{\bf v}'{\bf c}'_1{\bf c}'_2\rangle  = \nonumber\\
        &(E_0^{({\rm HF})} + \epsilon^{({\rm HF})}_{{\bf c}_1} + \epsilon^{({\rm HF})}_{{\bf c}_2} 
	- \epsilon^{({\rm HF})}_{\bf v}) 
\delta_{{\bf c}_1{\bf c}_1'} \delta_{{\bf c}_2{\bf c}_2'} 
\delta_{{\bf v}{\bf v}'} \nonumber\\
+ & (V_{{\bf c}_1{\bf c}_2,{\bf c}'_1{\bf c}'_2} - 
V_{{\bf c}_1{\bf c}_2,{\bf c}'_2{\bf c}'_1}) \delta_{{\bf v}{\bf v}'} 
     \nonumber\\
- & (V_{{\bf v}'{\bf c}_1,{\bf v}{\bf c}'_1} - 
     V_{{\bf v}'{\bf c}_1,{\bf c}'_1{\bf v}})
     \delta_{{\bf c}_2{\bf c}_2'} 
     \nonumber\\
- & (V_{{\bf v}'{\bf c}_2,{\bf v}{\bf c}'_2} - 
     V_{{\bf v}'{\bf c}_2,{\bf c}'_2{\bf v}})
     \delta_{{\bf c}_1{\bf c}_1'} 
~ .
\end{align}
The interpretation of these terms is straight-forward:
The first line of Eq.~(\ref{supp_eq22}) describes
independent motion of each particle in the system's band
structure, while the other terms describe the interaction
(direct and exchange) between the two electrons, and between the hole
and each of the electrons.

In accordance with Eqs. (\ref{supp_eq9}) and (\ref{supp_eq14}) we employ
an effective Hamiltonian for trion states as:
\begin{align}
\label{supp_eq28}
	\langle {\bf v}{\bf c}_1{\bf c}_2| \hat{H}^{(eeh)} 
	|{\bf v}'{\bf c}'_1{\bf c}'_2\rangle  & =
	(\epsilon_{{\bf c}_1} + \epsilon_{{\bf c}_2} 
	- \epsilon_{\bf v}) 
\delta_{{\bf c}_1{\bf c}_1'} \delta_{{\bf c}_2{\bf c}_2'} 
\delta_{{\bf v}{\bf v}'} \nonumber\\
& + (W_{{\bf c}_1{\bf c}_2,{\bf c}'_1{\bf c}'_2} - 
W_{{\bf c}_1{\bf c}_2,{\bf c}'_2{\bf c}'_1}) \delta_{{\bf v}{\bf v}'} 
     \nonumber\\
     & - (W_{{\bf v}'{\bf c}_1,{\bf v}{\bf c}'_1} - 
     V_{{\bf v}'{\bf c}_1,{\bf c}'_1{\bf v}})
     \delta_{{\bf c}_2{\bf c}_2'} 
     \nonumber\\
     & - (W_{{\bf v}'{\bf c}_2,{\bf v}{\bf c}'_2} - 
     V_{{\bf v}'{\bf c}_2,{\bf c}'_2{\bf v}})
     \delta_{{\bf c}_1{\bf c}_1'}
~ ,
\end{align}
again, keeping the general form as in Eq.~(\ref{supp_eq22}), but using $GW$ band-structure energies and a screened Coulomb interaction where necessary.

The Hamiltonian (\ref{supp_eq28}) also causes the formation of correlated
states $|T,{\bf K}\rangle$ ($\hat{=}$ trions) between the configurations (\ref{supp_eq21}), i.e.
\begin{equation}
\label{supp_eq23}
|T,{\bf K}\rangle = \sum_{{\bf v}{\bf c}_1{\bf c}_2} A^{(T,{\bf K})}_{{\bf v}{\bf c}_1{\bf c}_2} |{\bf v}{\bf c}_1{\bf c}_2\rangle
~ .
\end{equation}
Note that the summation is restricted to ${\bf c}_1 < {\bf c}_2$ (see above).
Here, ${\bf K}$ denotes the total momentum of the trion state (which, similar to excitons,
is a good quantum number following Bloch's theorem).
Since the wave numbers of the contributing configurations 
$({\bf v}{\bf c}_1{\bf c}_2)$ have to fulfill the
requirement ${\bf k}_v = {\bf k}_1 + {\bf k}_2 - {\bf K}$,
the sum in Eq.~(\ref{supp_eq23}) is reduced.

Similar to excitons and electron-electron states, a configuration (\ref{supp_eq21})
has a real-space amplitude
\begin{align}
\label{supp_eq24}
&\Phi_{{\bf v}{\bf c}_1{\bf c}_2}(x_h,x_1,x_2) =
\frac{1}{\sqrt{2}} \langle 0 | \hat{\psi}^\dagger(x_h) \hat{\psi}(x_1) \hat{\psi}(x_2) |{\bf v}{\bf c}_1{\bf c}_2\rangle\nonumber\\
&\qquad= \phi_{\bf v}^\ast(x_h) \frac{1}{\sqrt{2}} \left\lbrace \phi_{{\bf c}_1}(x_1) \phi_{{\bf c}_2}(x_2) - \phi_{{\bf c}_2}(x_1) \phi_{{\bf c}_1}(x_2) \right\rbrace
\end{align}
(with $x_h$, $x_1$ and $x_2$ being the coordinates of the hole and the two electrons)
and the corresponding linear combination for the trion state,
is given by
\begin{align}
\label{supp_eq24a}
&\Phi^{(T,{\bf K})}(x_h,x_1,x_2) = \sum_{{\bf v}{\bf c}_1{\bf c}_2} A^{(T,{\bf K})}_{{\bf v}{\bf c}_1{\bf c}_2}
\phi_{\bf v}^\ast(x_h)  \times\nonumber\\
&\qquad\qquad\times\frac{1}{\sqrt{2}} \left\lbrace \phi_{{\bf c}_1}(x_1) \phi_{{\bf c}_2}(x_2) - \phi_{{\bf c}_2}(x_1) \phi_{{\bf c}_1}(x_2) \right\rbrace
~ .
\end{align}

The expansion coefficients $A^{(T,{\bf K})}_{{\bf v}{\bf c}_1{\bf c}_2}$ and the energy 
$E^{(T,{\bf K})}$ of a trion $(T,{\bf K})$ result from the eigenvalue problem
\begin{equation}
\label{supp_eq25}
\sum_{{\bf v}'{\bf c}_1'{\bf c}_2'} 
\langle {\bf v}{\bf c}_1{\bf c}_2| \hat{H}^{(eeh)} |{\bf v}'{\bf c}'_1{\bf c}'_2\rangle A^{(T,{\bf K})}_{{\bf v}'{\bf c}'_1{\bf c}_2'}
= E^{(T,{\bf K})} A^{(T,{\bf K})}_{{\bf v}{\bf c}_1{\bf c}_2}
~ .
\end{equation}
Note that the energy $E^{(T,{\bf K})}$ is {\em not} a transition energy to occur
in an (optical) spectrum.
Such transitions (including absorption or emission of a photon) would occur
between the trion $|T,{\bf K}\rangle$ and an electron in the conduction band $|c{\bf K}\rangle$
(i.e. the single electron from which the absorption process starts, or the
single electron which remains after one electron and the hole of the trion have
recombined).
Note that the trion and the remaining electron must have the same wave number 
${\bf K}$ to
allow photon-assisted transitions between them.
Consequently, (optical) transition energies would be given by
\begin{equation}
\label{supp_eq26}
\Omega( |T,{\bf K}\rangle\leftrightarrow |c{\bf K}\rangle ) = 
E^{(T,{\bf K})} - \epsilon_{c{\bf K}}
~ .
\end{equation}
This is associated with a dipole moment
\begin{align}
\langle c{\bf K} | \hat{{\bf p}} |T,{\bf K}\rangle 
& = \sum_{{\bf v}{\bf c}_1{\bf c}_2} A^{(T,{\bf K})}_{{\bf v}{\bf c}_1{\bf c}_2} 
\langle c{\bf K} | \hat{{\bf p}} |{\bf v}{\bf c}_1{\bf c}_2\rangle\nonumber\\
& = \sum_{{\bf v}{\bf c}_1{\bf c}_2} A^{(T,{\bf K})}_{{\bf v}{\bf c}_1{\bf c}_2} 
\left( {\bf p}_{{\bf v}{\bf c}_1} \delta_{c{\bf K},{\bf c}_2} -
{\bf p}_{{\bf v}{\bf c}_2} \delta_{c{\bf K},{\bf c}_1} \right)
\label{supp_eq27}
~ .
\end{align}

\section{Hardware requirements for the calculations}\label{sec:req}
Due to the third particle, the configuration space is much larger for trions than for excitons.
A precise analysis of the required hardware is therefore necessary.

The number of occupied bands included in Eq.~(\ref{supp_eq23}) is denoted as $N_v$, the unoccupied bands as $N_c$ and the number of ${\bf k}$-points as $N_{\bf k}$.
The size of the trion Hamiltonian is $\sim \frac{1}{2} N_v \cdot N_c^2 \cdot N_{\bf k}^2$
(instead of $N_v \cdot N_c \cdot N_{\bf k}$ for excitons), including the factor $\sim$1/2
from the restriction that ${\bf c}_1$$<$${\bf c}_2$ (see Sec. \ref{subsec_eeh}).
Therefore the full Hamilton matrix can only be stored in memory for very small systems.

In Eq.~(\ref{supp_eq28}) all four terms
(kinetic energy, electron-electron interaction and electron-hole interaction)
include Kronecker deltas.
This allows to index these combinations properly 
and therefore to reduce the number of floating point operations
for every matrix-vector multiplication drastically,
in analogy to commonly known sparse matrix operations.
In addition, this makes it possible to restrict the memory requirement tremendously
and to share it efficiently between the processors.
The minimal required memory for non-zero elements of the Hamiltonian scales with
$N_c^2 \cdot (N_c^2 + N_v^2) \cdot N_{\bf k}^3$.

In the particular case of the (8,0) CNT, we usually employ
$32$ ${\bf k}$-points, 16 valence bands (out of a total of 128 valence bands, including spin)
and 16 conduction bands.
In these calculations, less than one percent of the Hamilton matrix elements is non-zero.
This causes memory requirements of about $33$\,GB.
When $64$ ${\bf k}$-points are employed, the requirement rises to about $262$\,GB.

In addition, up to ten vectors (of the dimension of the Hamiltonian) are required
for the iterative procedure of applying the Hamiltonian to states and 
for the evaluation of the spectrum (using, e.g., the Haydock method).
For the determination of trion states by an iterative diagonalization, on the other hand,
more vectors and therefore more memory is required.
For example, the determination of the lowest 5000 trion states (i.e., all states below
2.2\,eV in Fig.~1 of the main text) requires about $900$\,GB of memory,
again employing $32$ ${\bf k}$-points.

\section{Numerical results}
The trion calculations are carried out in the framework of many-body perturbation theory.
Therefore several steps are necessary, which will be discussed one after the other.

\subsection{DFT calculations}
\begin{figure}
  \includegraphics[width=.5\textwidth]{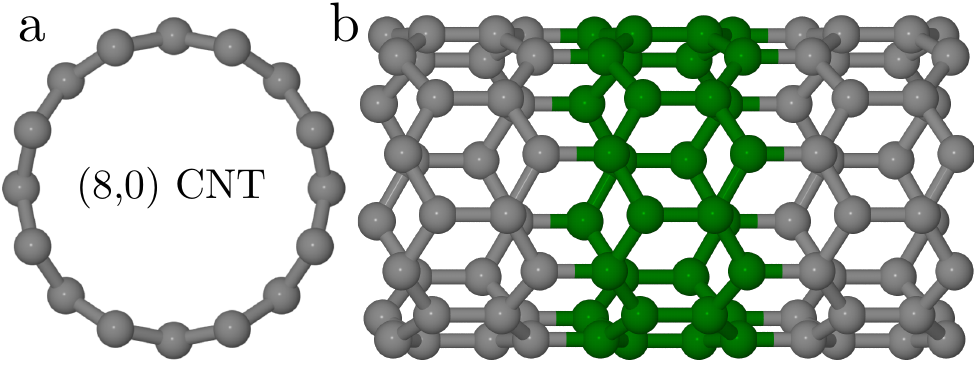}
\caption{
  Structure of the (8,0) CNT.
  (a) Front view and
  (b) side view.
  The unit cell is marked in green.
}\label{figStruc}
\end{figure}

As a basis for the many-body calculations of the (8,0) CNT we first carry out a DFT calculation
to provide an optimized geometry, DFT wave functions, and their band-structure
energies.
The $(8,0)$ CNT consists of $32$ carbon atoms in the unit cell,
the configuration is sketched in Fig.~\ref{figStruc}.
The structure has eight-fold rotational symmetry
along the tube axis,
two mirror symmetries are observed perpendicular to it
at the center and the border of the marked unit cell.
the local density approximation (in the parametrization of Perdew and Zunger \cite{lda_pz}).
Norm-conserving pseudopotentials \cite{pp_ham} in Kleinman-Bylander form \cite{pp_kb} are used.
We employ a mixed basis of 5 shells of Gaussian orbitals (38 functions 
with $s$, $p$, $d$, and $s^\ast$ symmetry, with decay constants 
from 0.1 to 3.0 $a_B^{-2}$) per carbon atom plus plane waves with a 
cutoff of $1$\,Ry.
The basis is unstable due to (almost) linear dependence among the basis 
functions.
Stability is recovered by eliminating those linear combinations which are
responsible for the instability, such that the resulting overlap matrix
is again positive definite.
A mesh of 16 ${\bf k}$-points in the one dimensional Brillouin zone is employed for the DFT.
We obtain an optimized lattice constant of  $4.22$\,\AA\ and a tube diameter of $6.29$\,\AA\  after full relaxation of the structure.
We employ a large unit cell perpendicular to the tube axis to guarantee a minimal distance of $10.6$\,\AA\ between the tubes.
With this we obtain the eigenvalues and wave functions which serve as input for the $GW$ calculation.

\subsection{$GW$ calculations}
The MBPT calculations are carried out within the $GW$ approximation for the
electron self energy.
This requires a second, auxiliary basis set in which two-point quantities
(e.g., the dielectric function, the screened Coulomb interaction etc.) 
are represented.
Again, a mixed basis of 4 shells of Gaussian orbitals (34 functions 
with $s$, $p$, $d$, $s^\ast$, and $f$ symmetry, with decay constants 
from 0.25 to 4.0 $a_B^{-2}$) per carbon atom plus plane waves with a 
cutoff of $1$\,Ry is used, which again has to be stabilized by elimination
of linear dependence.

The main purpose of the $GW$ calculation is to provide the band-structure
energies $\epsilon_{n,{\bf k}}$ required in Eqs. (\ref{supp_eq9}) and
(\ref{supp_eq28}), for the {\bf k}-points in which the exciton and trion 
states are represented.
The $GW$ self-energy operator further involves an integration in reciprocal 
space, which is carried out by a finite sum over {\bf q}-points. 
For reasons of consistency with the BSE and trion calculation (see below) 
we employ a grid of {\bf q}-points given by the differences between the
{\bf k}-points of the exciton (or trion) states.
Obviously this difference grid includes the $\Gamma$ point (${\bf q}=0$),
at which the Coulomb interaction diverges.
We solve this problem by replacing the screened Coulomb interaction
$W({\bf q}_i)$ by its average
$1/V_i \int_{V_i} W({\bf q}) d^3q$ in the reciprocal volume $V_i$ of
each grid point ${\bf q}_i$.

$GW$ calculations should include self-consistency in terms of the
resulting band-structure energies.
Here we simplify this requirement by including
a scissors shift of $0.1$\,Ry before the RPA screening and the $GW$ 
self-energy operator are evaluated.
Thereby we anticipate the opening of the gap
in a self-consistent approach.
Comparing to this, our procedure is accurate to within 0.05 eV.
Starting from a DFT-LDA gap of 0.58 eV, we finally obtain a $GW$ band gap
of 1.70 eV.
Similar results of $0.6$ eV (DFT) and $1.75$\,eV ($GW$) have been found previously \cite{CNT_Louie}.

\subsection{BSE calculations / optical absorption and luminescence spectra}
\begin{figure}
  \includegraphics[width=.5\textwidth]{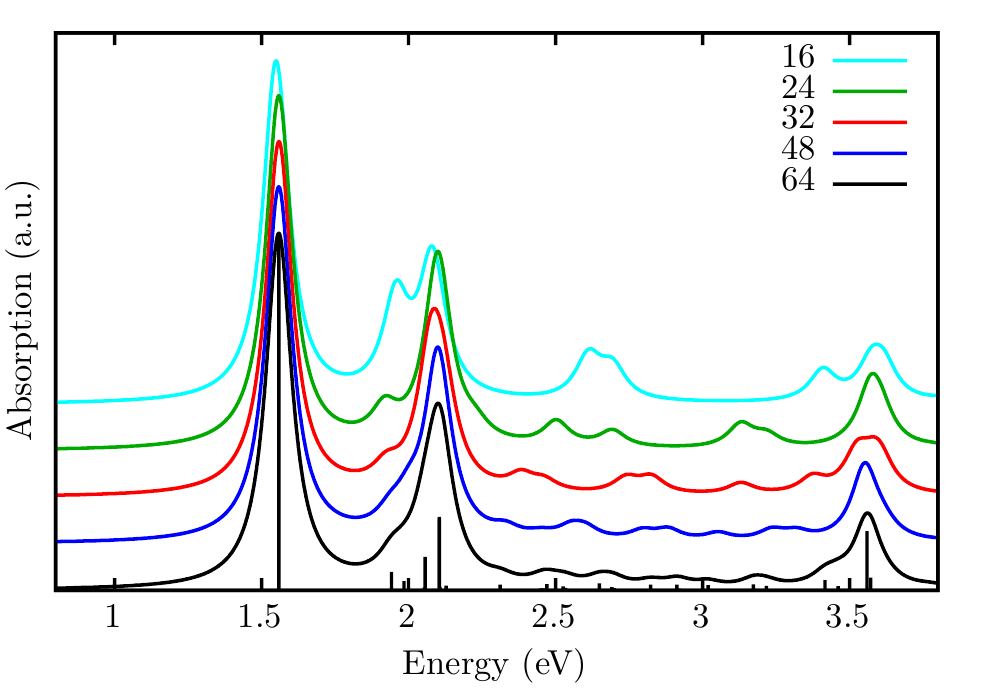}
\caption{
Exciton absorption spectrum.
The spectra include different vertical offset to improve visibility.
An artificial broadening of $0.054$\,eV is employed.
}\label{figS1}
\end{figure}
The optical properties are then obtained by solving the BSE.
We have tested convergence with respect to the {\bf k}-points, using
grids from 16 to 64 points.
The resulting optical spectra are compiled in Fig.~\ref{figS1}.
We observe results converged better than $0.1$\,eV even for meshes with only 16 ${\bf k}$-points.
Especially the first optical peak at $1.56$\,eV is very stable,
as well as the first dark exciton at $1.090$\,eV (not shown).
Both this optical peak and the next one at $1.94$\,eV are in good agreement with previous studies,
as well as with experimental data (see Ref.~\cite{CNT_Louie,Chang2004,CNT_GdW,CNT_impurity}).

The calculation of luminescence spectra
is analogous as explained in the next section (Eq.~(\ref{supp_eq29})).

\subsection{Trion calculations / optical absorption and luminescence spectra}
\begin{figure}
  \includegraphics[width=.5\textwidth]{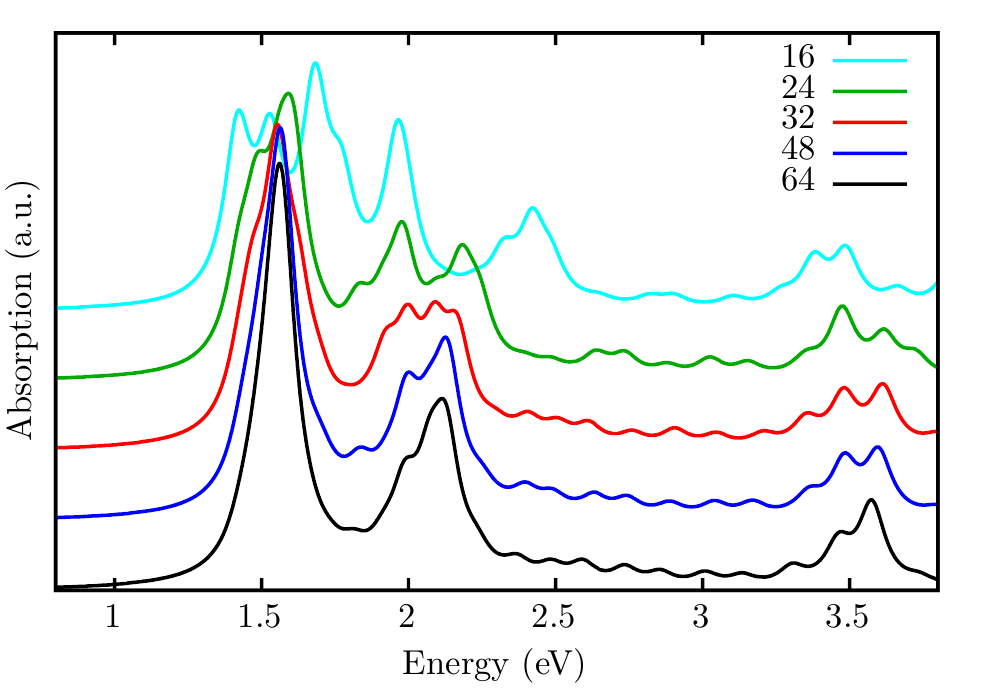}
\caption{
Trion absorption spectrum
obtained using the Haydock recursion method.
The spectra include different vertical offset to improve visibility.
An artificial broadening of $0.054$\,eV is used.
}\label{figS2}
\end{figure}

In the last step we include an additional electron, and consider trion states
as linear combinations of three-particle configurations.
As described in Sec.~\ref{sec:req}, the direct diagonalization of the 
trion Hamilton operator is not possible due to its high dimension 
($\sim$10$^6$), which depends on the number of bands and, in particular, 
on the number of {\bf k}-points to be considered.
For the evaluation of the optical spectrum the Haydock recursion 
method \cite{haydock} and real-time integration followed by Fourier 
transform \cite{timeint} are a well established techniques.
For the determination of a limited number of eigenstates at low energy,
we employ iterative diagonalization techniques \cite{SLEPc1,SLEPc2}.
We have carefully checked the convergence behaviour of the spectra
(using the Haydock method and using Fourier transform after real-time 
integration) and of the eigenstates with respect to the
{\bf k}-point sampling (see Fig. \ref{figS2}).
We have further checked that in the low-energy part of the spectrum
(accessible by iterative diagonalization) the eigenstates and the
Haydock method yield the same spectral data.

To converge the full trion spectrum a larger number of ${\bf k}$-points 
has to be used than for the excitons.
Fortunately, the low-energy results from 32 ${\bf k}$-point are already reasonably 
well converged  (in comparison with data from larger sets, as shown 
in Fig.~\ref{figS2}).
In analogy to the dipole moment for an exciton (see Eq.~(\ref{supp_eq7b})) 
the optical dipole moment for a trion are calculated by Eq.~(\ref{supp_eq27}).
Note that in contrast to excitons, the initial (or final) state from 
(or to) which a transition occurs is {\em not} the ground state $|0\rangle$
but a single-electron state in a conduction band, $|c{\bf K}\rangle$.
Due to momentum conservation in the optical transition, this electron's
momentum must equal the trion momentum, {\bf K}.
For Fig. \ref{figS2} we have chosen ${\bf K}=0$, and the single-electron
initial/final state is the CBM.

\begin{figure}
  \includegraphics[width=.4\textwidth]{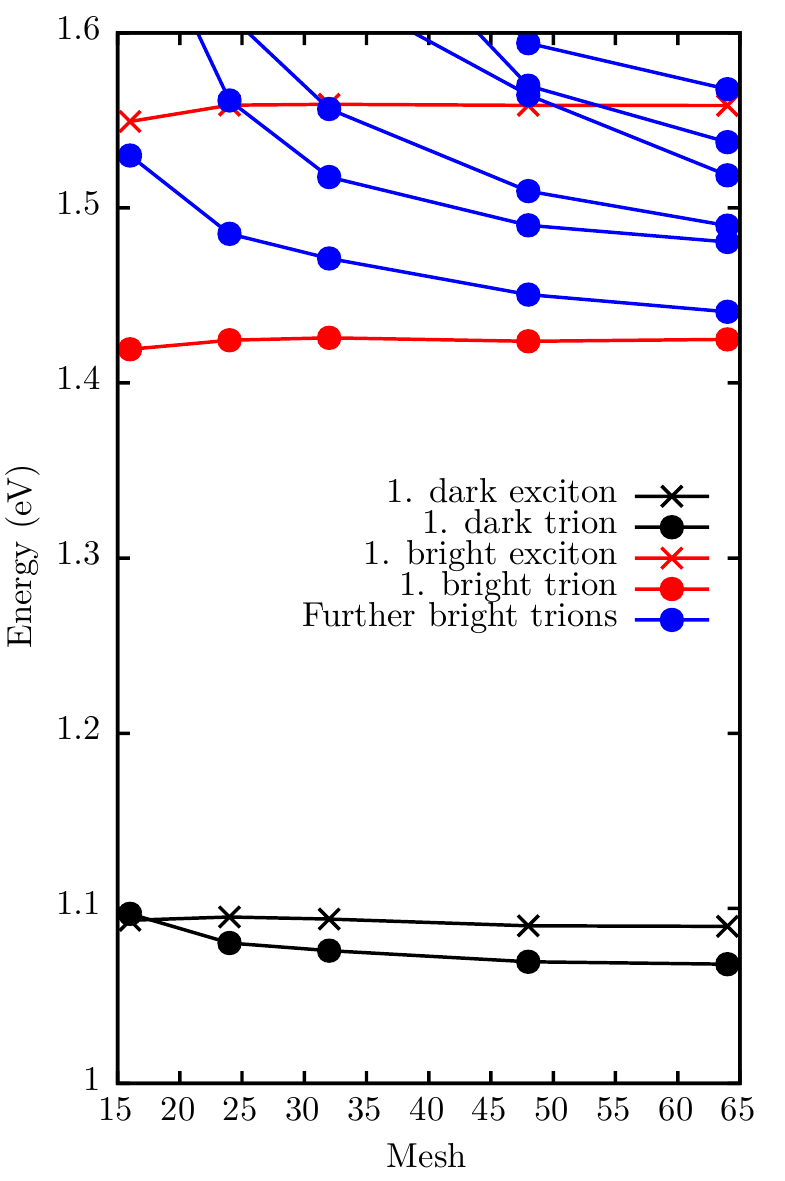}
\caption{
Convergence of the trion states with respect to the number of ${\bf k}$-points.
The first dark (bright) exciton and trion are marked by black (red) crosses and dots.
In addition, further trions (that are found in the manageable number of lowest states) are indicated by blue dots.
}\label{figS3}
\end{figure}

Although the spectra from various {\bf k}-point grids are in reasonable agreement with each other,
their exact shape slightly depends on the ${\bf k}$-point mesh, in particular
at higher energy.
Fortunately, the behaviour of the low-energy states is very stable.
Fig.~\ref{figS3} shows the convergence of two excitons and of the related
trions with respect to the grid density.
In particular the energies of the lowest-energy trion (at 1.1 eV) 
and of the lowest dipole-allowed trion (at 1.4 eV) are very stable,
similar to the excitons.

Optical absorption due to trions starts from a single-electron state
and ends in a trion state.
In contrast, luminescence (see Fig.~1 b of the main text) starts from
the trions and ends in a single-electron state $|c{\bf K}\rangle$, 
again conserving total momentum.
In this context several issues have to be considered.
First, after the original excitation process the trions
will (after some time) equilibrate according to the system's temperature.
By including a temperature-dependent Boltzmann distribution, we
obtain a luminescence spectrum
\begin{align}
  \nonumber L(E) \propto  \sum_{c,{\bf K},|T,{\bf K}\rangle} \text{e}^{-\beta E^{(T,{\bf K})}}
   &|\langle c{\bf K} | \hat{{\bf p}} | T,{\bf K} \rangle|^2 \times\\
  \times \delta(E-&\Omega( |T,{\bf K}\rangle\leftrightarrow |c{\bf K}\rangle ) )
\label{supp_eq29}
\end{align}
with $\beta = 1/(k_BT)$.
Secondly, this expression requires to consider all possible momenta
{\bf K} and final-state band indices $c$.
However, we find in our calculations that the dipole moments are
particularly strong near ${\bf K}=0$ and quickly decrease for ${\bf K}\ne 0$.
Furthermore, 
$E^{(T,{\bf K})}$ increases for ${\bf K}\ne0$, leading to a decreasing
Boltzmann factor and decreasing contribution to Eq. (\ref{supp_eq29}).
We have found that focusing on ${\bf K}=0$ is sufficient, since the inclusion
of other momenta has no recognizable effect on the luminescence spectrum
at room temperature.
It should also be noted that the optically allowed trions near 1.4\,eV show 
strong dipoles only for $c=\text{CBM}$ (i.e., the remaining electron in the lowest 
conduction band) and much smaller contribution from other bands.
Transitions to higher bands $c$ occur at lower energy, but are dipole forbidden in the present case. 
\vspace*{3cm}

%

\end{document}